\documentclass[a4paper,onecolumn,10pt,accepted=2025-12-21]{quantumarticle}
\pdfoutput=1
\usepackage[utf8]{inputenc}
\usepackage[english]{babel}
\usepackage[T1]{fontenc}
\usepackage{amsmath}
\usepackage{amssymb}
\usepackage{multirow}
\usepackage{braket}
\usepackage{ dsfont }
\usepackage{amsthm}
\usepackage{soul}
\usepackage{tikz}
\usepackage{lipsum}
\usepackage[numbers,sort&compress]{natbib}
\usepackage{stmaryrd}
\usepackage{lipsum}

\usepackage{graphicx}

\usepackage{mathtools}
\usepackage{pifont}

\definecolor{amethyst}{rgb}{0.6, 0.4, 0.8}
\definecolor{quantumpurple1}{RGB}{142, 68, 173}  
\definecolor{quantumpurple4}{RGB}{224, 202, 240} 
\definecolor{quantumpurple2}{RGB}{192, 150, 219}
\definecolor{quantumpurple3}{RGB}{188, 54, 118}  
\definecolor{bluegray}{RGB}{136,141,186}

\usepackage[colorlinks, bookmarksopen, bookmarksnumbered, citecolor=amethyst, linkcolor=amethyst, urlcolor=amethyst]{hyperref}

\usepackage{ifthen}

\newcommand{\Andrea}[2][]{%
  \textcolor{orange}{#2}%
  \ifthenelse{\equal{#1}{}}{}{%
    \textcolor{orange}{\textbf{[Andrea: #1]}}%
  }%
}

\usepackage{orcidlink}
\usepackage{tikz,ifthen}
\usepackage{bbold}
\usepackage{tikz-network}
\usetikzlibrary{patterns,decorations.pathreplacing,calligraphy}
\usetikzlibrary{shapes,arrows.meta,decorations.pathmorphing}

\newcommand{\rrangle}{\rangle\!\rangle}
\newcommand{\llangle}{\langle\!\langle}
\newcommand{\kket}[1]{|#1\rrangle}
\newcommand{\bbra}[1]{\llangle #1|}

\newcommand{\titleinfo}{{Quantum Complexity and Chaos in Many-Qudit Doped Clifford Circuits}}

\begin{document}
\title{\titleinfo} 

\author{Beatrice Magni~\orcidlink{0009-0009-8577-0525}}
\email{bmagni@uni-koeln.de}
\affiliation{Institut für Theoretische Physik, Universität zu Köln, Zülpicher Strasse 77, 50937 Köln, Germany}

\author{Xhek Turkeshi~\orcidlink{0000-0003-1093-3771}}
\email{xturkesh@uni-koeln.de}
\affiliation{Institut für Theoretische Physik, Universität zu Köln, Zülpicher Strasse 77, 50937 Köln, Germany}

\begin{abstract}
We investigate the emergence of quantum complexity and chaos in doped Clifford circuits acting on qudits of odd prime dimension $d$.
Using doped Clifford Weingarten calculus and a replica tensor network formalism, we derive exact results and perform large-scale simulations in regimes challenging for tensor network and Pauli-based methods.
We begin by analyzing generalized stabilizer entropies—computable magic monotones in many-qudit systems—and identify a dynamical phase transition in the doping rate, marking the breakdown of classical simulability and the onset of Haar-random behavior. The critical behavior is governed by the qudit dimension and the magic content of the non-Clifford gate.
Using the qudit $T$-gate as a benchmark, we show that higher-dimensional qudits converge faster to Haar-typical stabilizer entropies. For qutrits ($d=3$), analytical predictions match numerics on brickwork circuits, showing that locality plays a limited role in magic spreading.
We also examine anticoncentration and entanglement growth, showing that $\mathcal{O}(\log N)$ non-Clifford gates suffice for approximating Haar expectation values to precision $\varepsilon$, and relate antiflatness measures to stabilizer entropies in qutrit systems.
Finally, we analyze out-of-time-order correlators and show that a finite density of non-Clifford gates is needed to induce chaos, with a sharp transition fixed by the local dimension—twice that of the magic transition.
Altogether, these results establish a unified framework for diagnosing complexity in doped Clifford circuits and deepen our understanding of resource theories in multiqudit systems.
\end{abstract}

\maketitle

\section{Introduction}
In recent years, experimental quantum platforms have evolved into controllable many-body systems, enabling the realization of complex out-of-equilibrium dynamics~\cite{preskill2018quantum}. Within this setting, quantum circuits provide a versatile framework for generating and probing dynamical phases~\cite{fisher2023random}---collectively referred to as synthetic quantum matter~\cite{grass2025colloquium}. 
These phases typically lack conventional order parameters, such as magnetization, and are more naturally characterized through quantum information-theoretic diagnostics.
Complexity, in this context, serves as a central organizing principle: it delineates phases by the classical difficulty of simulating their dynamics, ranging from efficiently simulable regimes, such as matrix product states~\cite{schollwock2011the}, to classically intractable ones, such as chaotic quantum dynamics~\cite{abanin2019colloquium,sierant2025manybody}.

Among various indicators, quantum magic has emerged as a fundamental metric---particularly relevant for fault-tolerant quantum computation based on Clifford operations with magic-state inputs~\cite{gottesman1997stabilizer,gottesman1998theory,bravyi2005universal,Turkeshi2024}. 
In this regard, Clifford gates form a key component of quantum error correction and allow for the efficient preparation of a broad class of quantum states, including highly entangled ones~\cite{nahum2017quantum,sierant2023entanglement}.
Yet, despite their versatility, Clifford circuits are classically simulable and therefore insufficient to achieve quantum advantage~\cite{aaronson2004improved,Bravyi2016,Bravyi2019simulationofquantum}. 
Quantum magic—or nonstabilizerness—is required to overcome this limitation.
This resource is typically introduced through non-Clifford operations such as $T$-gates~\cite{chitambar2019quantum,liu2022manybody,leone2022stabilizer}. 
Architectures where a finite and controlled number of non-Clifford gates are injected into otherwise stabilizer dynamics are known as doped Clifford circuits~\cite{leone2024learning,bejan2024dynamical,lami2024quantum,lami2024unveiling,mello2024hybrid,haug2024probingquantumcomplexityuniversal,fux2025disentanglingunitarydynamicsclassically,nakhl2025stabilizer,liu2024classicalsimulabilitycliffordtcircuits,magni2025anticoncentrationcliffordcircuitsbeyond,leone2025noncliffordcostrandomunitaries,turkeshi2025magicspreadingrandomquantum, Mao_2025, aditya2025mpembaeffectsquantumcomplexity,paviglianiti2025emergencegenericentanglementstructure,lóio2025quantumstatedesignsmagic,aditya2025growthspreadingquantumresources,magni2025anticoncentrationstatedesigndoped,varikuti2025impactcliffordoperationsnonstabilizing,santra2025complexitytransitionschaoticquantum,haferkamp2022, Heinrich2019robustnessofmagic, True2022transitionsin}.
Recent studies have shown that even a single $T$-gate can dramatically alter the scaling properties of Clifford circuits, e.g., shifting the entanglement spectrum toward that of Haar-random unitaries~\cite{Zhou_2020} or obtaining $\epsilon$-approximate designs when applied to random stabilizer states~\cite{haferkamp2022}. 
Moreover, stabilizer entropies provide a scalable measure of nonstabilizerness~\cite{leone2022stabilizer, haug2023stabilizer, haug2025efficientwitnessingtestingmagic, lami2023nonstabilizerness, hoshino2025stabilizerrenyientropyconformal,hou2025stabilizerentanglementmagichighway}, revealing a linear growth of magic with the number of injected gates, followed by saturation at universal values~\cite{haug2024probingquantumcomplexityuniversal,szombathy2025spectralpropertiesversusmagic, szombathy2025independentstabilizerrenyientropy,bejain2024,fux2025disentanglingunitarydynamicsclassically,liu2024classicalsimulabilitycliffordtcircuits,masotlima2024}.
Complementary diagnostics, such as out-of-time-ordered correlators (OTOCs)~\cite{Yoshida_2019}, indicate that a doped Clifford circuit must include $\Theta(N)$ non-Clifford gates to faithfully reproduce signatures of quantum chaos~\cite{Leone2021quantumchaosis}. 
While these results are well established for qubit systems~\cite{bittel2025completetheorycliffordcommutant}, far less is known in the qudit case, where the structure of the Clifford group and its design properties fundamentally differ~\cite{gross2021schurweyl,turkeshi2025magicspreadingrandomquantum,magni2025anticoncentrationcliffordcircuitsbeyond}--- a problem that is particularly relevant for platforms based on multilevel qudits. 

This work investigates the emergence of quantum complexity in doped Clifford circuits acting on qudit systems with odd prime dimension as a function of the number of non-Clifford gates $N_T$ injected. 
Our analysis is guided by several diagnostic indicators. First, we consider generalized stabilizer entropies (GSEs)~\cite{turkeshi2025magicspreadingrandomquantum}--- a family of nonstabilizerness measures extending stabilizer entropies to systems with higher local dimension. 
We track the system’s evolution with the doping rate $q = N_T / N$ and identify a dynamical phase transition at a critical threshold $q_c(d)$. Below this point, the dynamics remain non-chaotic, and the magic density is nonmaximal. Above $q_c$, the system rapidly enters a regime characterized by universal magic scaling. Higher-dimensional qudits reach this regime faster and exhibit larger asymptotic values, growing logarithmically with $d$.
We test these analytical predictions using brickwork Clifford circuits for qutrit systems ($d = 3$), where non-Clifford resources are periodically injected at fixed positions.
Surprisingly, despite the locality of the circuit architecture, the behavior predicted for globally doped Clifford circuits quantitatively captures the dynamics of the local setting.
This result highlights the limited role of locality in the spreading of magic across many-body systems, in sharp contrast with the ballistic propagation of entanglement entropy or of operator magic in the Heisenberg picture~\cite{dowling2024magic,dowling2025bridgingentanglementmagicresources}.
Complementarily, we study anticoncentration and entanglement properties in doped Clifford circuits, quantitatively captured by the inverse participation ratios and Rènyi purities, respectively.
We derive exact expressions for both as functions of the doping rate, finding that $N_T = O(\log N)$ suffices to approximate with precision $\varepsilon$ the Haar unitary value. In the qutrit case, we establish a direct connection to antiflatness measures~\cite{tirrito2024quantifyingnon,turkeshi2023measuring}.
Finally, we examine the onset of quantum chaos using out-of-time-ordered correlators (OTOCs).
We demonstrate that achieving universal behavior characteristic of Haar-random circuits requires a number of dopings scaling linearly with system size, $N_T\sim N$, corresponding to a finite density $q= N_T/N=O(1)$. Moreover, we find that the critical doping threshold decreases with increasing local dimension, with the transition to chaotic dynamics occurring at a density that is approximately twice the critical magic threshold, $q_c(d)$. These findings reinforce the pivotal role of the local dimension, along with the doping injection, in governing the growth of complexity in qudit doped circuits and reveal the presence of an intermediate regime characterized by maximal magic in the absence of full chaoticity, revealed by the OTOCs.
Altogether, these findings establish a unified framework for diagnosing complexity in doped Clifford circuits beyond qubits and deepen our understanding of resource theories in multiqudit architectures.

\section{Methods}
This section sets the formalism and methodologies underlying our work. 
After briefly presenting Pauli and Clifford group, and stabilizer states, we present the central technique at the basis of our results: the Clifford Weingarten calculus.
This dictionary maps Clifford group moments to contractions in replica space, allowing quantum circuits to be represented as replica tensor network states with tailored boundary conditions.

\paragraph{Preliminaries.}
Throughout this work, we consider a system of $N$ qudits, each with local Hilbert space $\mathcal{H}_d \cong \mathbb{C}^{d}$, where $d$ is an odd prime. The total Hilbert space is $\mathcal{H} = \mathcal{H}_d^{\otimes N}$, with total dimension $D = d^N$. 
Given the $d$-th root of unity $\omega = \exp[2\pi i/d]$, the generalized Pauli operators $X$ and $Z$ are defined as
\begin{equation}
    X = \sum_{m=0}^{d-1} |m \oplus 1\rangle\langle m| \;, \qquad Z = \sum_{m=0}^{d-1} \omega^m |m\rangle\langle m| \;,
\end{equation}
where $\oplus$ denotes addition modulo $d$ in the finite field $\mathbb{Z}_d$. 
These elementary operators generate the Pauli group, defined as the set of tensor products of Pauli operators with a phase $\tilde{\mathcal{P}}_N \equiv \lbrace \omega^\phi \bigotimes_{j=1}^{N} (X^{a_j} Z^{b_j}) \; \Big| \; \phi, a_j, b_j \in \mathbb{Z}_d \rbrace$. 
We also define the \textit{unsigned} Pauli group $\mathcal{P}_N \equiv \tilde{\mathcal{P}}_N / \mathrm{U}(1)$, which identifies Pauli strings up to global phases.

The Clifford group $\mathcal{C}_N \subset \mathrm{U}(D)$ is the subgroup of unitary operators that normalize the Pauli group, meaning that for any $P \in \tilde{\mathcal{P}}_N$, a Clifford unitary $C \in \mathcal{C}_N$ maps $P$ to another Pauli string $CPC^\dagger \in \tilde{\mathcal{P}}_N$. 
We define the set of stabilizer states $\mathrm{STAB}_{d,N}$ as the set of vectors in Hilbert space obtained by applying a Clifford unitary $C \in \mathcal{C}_N$ to the reference state $|0\rangle$, namely
\begin{equation}
    \mathrm{STAB}_{d,N} = \{ C|0\rangle \;| \;C\in \mathcal{C}_N\}\;.
\end{equation}
Clifford gates and measurements of Pauli strings are efficiently implementable when acting on stabilizer states via the Gottesmann-Knill theorem~\cite{aaronson2004improved,Heinrich2019robustnessofmagic}. 

Clifford circuits and stabilizer states form the backbone of quantum error correction and fault-tolerant quantum computation~\cite{nielsen2010quantum}.
Nevertheless, they are not sufficient for universal quantum computing: additional non-Clifford resources are required to achieve universality. 
These are commonly characterized through \emph{magic states}, which quantify the non-Clifford content required to prepare a given state~\cite{bravyi2005universal}.  
A widely used strategy to introduce such resources is \emph{magic state injection}, where a non-Clifford gate is implemented by preparing a magic ancilla and teleporting it into the system via a \emph{gadget circuit}~\cite{gottesmann1999demonstrating,Bravyi2016}. This protocol identifies a class of non-Clifford gates that can be implemented fault-tolerantly, closely related to the structure of the so-called Clifford hierarchy~\cite{howard2012qudit,desilva2025cliffordhierarchyqubitqudit}. In the following, we focus on gates within these classes, and in particular, on quantum circuits composed of Clifford operations interspersed with such non-Clifford gates-- the so-called doped Clifford circuits--as a model for controllable non-Clifford dynamics.

\paragraph{Replica tensor networks and Clifford Weingarten calculus.}  
Before introducing our setup and main results, we briefly outline the core techniques underpinning our analysis: the replica formalism and the Clifford Weingarten calculus.

We make use of the Choi isomorphism, which maps operators $A$ acting on $\mathcal{H}$ to vectors $|A\rrangle = (A \otimes I)|\Phi\rangle$ in $\mathcal{H}^{\otimes 2}$, where $|\Phi\rangle = \sum_{i=0}^{D-1} |i\rangle \otimes |i\rangle / \sqrt{D}$ is the so-called Choi state~\cite{mele2024introduction}.
Under this isomorphism, inner products rephrase to overlaps $\mathrm{tr}(A^\dagger B)=\llangle A|B\rrangle$, and the adjoint action of an operator $K$ to matrix multiplications $K A K^\dagger|\Phi\rangle = K\otimes K^* |A\rrangle$. 

Recall that a quantum circuit consists of local gates $U_\lambda \in \mathrm{U}(D)$ acting on subsets of sites $\lambda \subset \{1, 2, \dots, N\}$.
By locality, we mean that each gate $U_\lambda$ acts non-trivially only on the sites in $\lambda$, leaving the other untouched\footnote{ 
Formally, these gates are obtained by embedding $U \in \mathrm{U}(d^{|\lambda|})$ into the larger space via a permutation operation
$U_\lambda = S_\sigma \left(U \otimes I^{\otimes (N - |\lambda|)} \right) S_\sigma^\dagger$.
The permutation $\sigma \in \mathrm{S}_N$ maps $\{1,2,\dots,|\lambda|\} \subset \{1,\dots,N\}$ to the target sites $\lambda$ and is represented on $\mathcal{H}$ via $S_\sigma |i_1, \dots, i_N\rangle = |i_{\sigma^{-1}(1)}, \dots, i_{\sigma^{-1}(N)}\rangle$.}.  
The full circuit is then specified by a geometry $\mathfrak{A} = \{\Lambda_s \:|\: s = 0, \dots, t\}$, where $t$ is the circuit depth and each $\Lambda_s$ is a set of disjoint subsets $\lambda$ indicating the support of the gates applied at layer $s$. 

This remark, together with the fact that quantum complexity is encoded in non-linear features of quantum states, implies that the fundamental objects we need to evaluate are the $k$-replica overlaps
\begin{equation}
\mathcal{O}_{A,B} \equiv \mathbb{E}_{C_\lambda} \left[\llangle {A^\dagger} |(C_\lambda \otimes C_\lambda^*)^{\otimes k} |B \rrangle\right],
\end{equation}
where $A$, $B$ are operators acting on the $k$-replica system $\mathcal{H}^{\otimes k}$, and $C_\lambda$ are random Clifford gates. 
The computation of $\mathcal{O}_{A,B}$ leverages the so-called Clifford Weingarten calculus. 
By the linearity of expectation value, the fundamental component of this calculation is the \textit{replica transfer matrix}, or moment operator~\cite{magni2025anticoncentrationcliffordcircuitsbeyond,sauliere2025universalityanticoncentrationchaoticquantum}
\begin{equation}
\mathcal{T}_\lambda = \mathbb{E}_{C_\lambda} [(C_\lambda \otimes C_\lambda^*)^{\otimes k}].
\end{equation}

By Schur-Weyl duality~\cite{gross2021schurweyl}, the replica transfer matrix is naturally described in terms of the $k$-th \textit{Clifford commutant}, defined as the set of all operators $A$ acting on $\mathcal{H}^{\otimes k}$ that commute with all $k$-fold tensor powers of Clifford unitaries, namely $[A, C_\lambda^{\otimes k}] = 0$ for any $C_\lambda \in \mathcal{C}_{N}$. 
Since the Clifford group $\mathcal{C}_{N}$ is a subgroup of the full unitary group $\mathrm{U}(D)$, its commutant contains that of the unitary group $\mathrm{Comm}_k(\mathcal{C}_N) \supseteq \mathrm{Comm}_k(\mathrm{U}(D))$, where $\mathrm{Comm}_k(\mathrm{U}(D))$ is spanned by the action of the symmetric group $\mathrm{S}_k$, represented as permutations of the $k$ replicas on $\mathcal{H}^{\otimes k}$~\cite{mele2024introduction}.

To fully characterize the larger commutant $\mathrm{Comm}_k(\mathcal{C}_N)$, we invoke its correspondence with the \textit{stochastic Lagrangian subspaces} $\sigma \subseteq \mathbb{Z}_d^{2k}$. These subspaces are defined by the following properties: (i) for every $(\mathbf{x}, \mathbf{y}) \in \sigma$, $ \sum_{i=0}^{k-1} (x_i^2 - y_i^2) \equiv 0 \mod d$, (ii) $\sigma$ is a $k$-dimensional subspace, (iii) the all-ones vector $(1, 1, \dots, 1)$ belongs to $\sigma$. 
The set of all such subspaces is denoted as $\Sigma_k(d)$, whose cardinality is given by
\begin{equation}
|\Sigma_k(d)| = \prod_{m=0}^{k-2} (d^m + 1) = (-1; d)_{k-1},
\label{eq:comm_dim}
\end{equation}
where $(a; \xi)_n = \prod_{k=0}^{n-1} (1 - a \xi^k)$ is the $q$-Pochhammer symbol~\cite{Kac2002}.

Each stochastic Lagrangian subspace $\sigma$ corresponds to an operator $T_\sigma \in \mathrm{Comm}_k(\mathcal{C}_N)$ acting on the replica space of each qudit $\mathcal{H}_d^{\otimes k}$, i.e., $T_\sigma|x_1,\dots,x_k\rangle=|x_{\sigma^{-1}(1)},\dots,x_{\sigma^{-1}(N)}\rangle$.
For example, when $\sigma\in \mathrm{S}_k$, these are permutation operators among the replicas\footnote{To distinguish between operators acting on individual qudits across $k$ replicas and permutations acting within a single replica over $N$ qudits, we denote them by $T_\sigma$ and $S_\sigma$, respectively.}. 
For notational convention, the Choi vectors associated with these replica operators are denoted $|\sigma\rrangle\equiv |T_\sigma\rrangle $. 
Denoting $r=|\lambda|$, these vectors enable the construction of a representation for $k$ copies of $r$-many qudits as
\begin{equation}
|\sigma\rrangle_\lambda = \bigotimes_{i \in \lambda} |\sigma_i\rrangle\;.
\end{equation}
For $r\geq k-1$, the vectors $|\sigma\rrangle_\lambda$ form a linearly independent set, ensuring that $|\mathrm{Comm}_k(\mathcal{C}_{r})| = |\Sigma_k(d)|$. Otherwise, linear dependencies arise among these vectors. 
The \textit{Schur-Weyl duality} for Clifford unitaries allows us to express the replica transfer matrix in terms of the commutant elements as~\cite{gross2021schurweyl,turkeshi2025magicspreadingrandomquantum,magni2025anticoncentrationcliffordcircuitsbeyond,bittel2025completetheorycliffordcommutant,leone2025noncliffordcostrandomunitaries}
\begin{equation}
\mathcal{T}_\lambda = \sum_{\pi, \sigma \in \Sigma_k(d)} \mathrm{Wg}_{\pi, \tau}\left(d^r\right) \kket{\pi}_\lambda \bbra{\sigma}_\lambda.
\end{equation}
Here, $\mathrm{Wg}_{\pi, \tau}\left(d^r\right)$ is the pseudo-inverse of the Gram matrix 
$\mathrm{G}_{\sigma, \pi} = \prod_{i\in \lambda} \llangle \sigma_i | \pi_i \rrangle.$
Graphically, we can represent the transfer matrix  on $r$ qudits as 
\begin{equation}
\mathbb{E}_\mathrm{Haar}\left[
\begin{tikzpicture}[baseline=(current bounding box.center), scale=0.4]
    \draw [decorate,decoration={brace}] (-3,+1.5) -- node[above]{$k$}(-1,3.5);
    \foreach \k in {8,7,...,1} {
        \pgfmathparse{mod(\k,2)==1?"quantumpurple1":"quantumpurple2"}
        \edef\col{\pgfmathresult}
        \foreach \i in {1,...,6} {
            \draw[thick] (-3.5+\i+0.2*\k,1.5+0.2*\k) -- (-3.5+\i+0.2*\k,-1.5+0.2*\k);
        }
        \draw[thick, fill=\col, rounded corners=3pt] (-3+0.2*\k,-1+0.25*\k) rectangle (3+0.2*\k,1+0.2*\k);
    }
    \draw [decorate,decoration={brace,mirror}] (-3,-1.5) -- node[below]{$r$}(4,-1.5);

\end{tikzpicture}
\right] = \sum_{\sigma,\pi\in \Sigma_k(d)} 
\begin{tikzpicture}[baseline=(current bounding box.center), scale=0.4]
    \foreach \i in {1,...,6} {
        \draw[ultra thick] (-3.5+\i,1.5) -- (-3.5+\i,-1.5);
        \node at (-3.5+\i,1.9) {$\pi$};
        \node at (-3.5+\i,-1.9) {$\sigma$};
    }
    \draw[ultra thick, fill=quantumpurple3, rounded corners=3pt] (-3,-1) rectangle (3,1);
    \draw [decorate,decoration={brace,mirror}] (-3,-2.5) -- node[below]{$r$}(3.,-2.5);
\end{tikzpicture}\label{fig:transfer_matrix}
\end{equation}
and the Gram matrix arising from contracting commutant vectors as
\begin{equation}
\begin{tikzpicture}[baseline={(current bounding box.center)}, thick]
  \node at (-1.2,0) {$\pi$};

  \draw[line width=1.2pt] (-1,0) -- (0.5,0);

  \node at (-0.25,0) { $\times$};
  \node at (0.75,0) {$\sigma$};
  \node at (1.15,0) {$=$};
\node at (1.5,0) {$\pi$};

  \draw[black, line width=1.2pt] (1.75,0) -- (2.2,0);
  \draw[black, line width=1.2pt] (2.8,0) -- (3.25,0);
  \filldraw[quantumpurple4, draw=black, line width=1.2pt] (2.5,0) circle (0.3);
  \node at (3.5,0) {$\sigma$};

  \node at (2.5,0) {$\mathrm{G}$};
\end{tikzpicture}\;.
\label{fig:gram}
\end{equation}
A fundamental property of the Gram and Weingarten matrices is that their marginal sum is constant, namely
\begin{equation}
\sum_{\pi\in \Sigma_k(d)} \mathrm{G}_{\pi, \sigma}(d^r) = \left(\sum_{\pi\in \Sigma_k(d)} \mathrm{Wg}_{\pi, \sigma}(d^n)\right)^{-1}=\mathcal{G}_{d,k,r}\equiv d^{kr} (-d^{-r}; d)_{k-1}.\label{eq:weing_sum}
\end{equation} 
Between random Clifford gates, doped Clifford circuits include the insertion of non-Clifford magic gates $K$. Upon averaging, the contractions between commutant vectors involve the operator $\mathcal{K} \equiv K \otimes K^*$, leading to the so-called \emph{doped Gram matrix}, defined as $
\tilde{\mathrm{G}}_{\sigma \pi} = \llangle \sigma | \mathcal{K}^{\otimes k} | \pi \rrangle$ and  graphically represented by
\begin{equation}   
\begin{tikzpicture}[baseline={(current bounding box.center)}, thick]
  \node at (-1.,0) {$\pi$};

  \node[draw=black, anchor=center] (box) at (0,0) {$\mathcal{K}^{\otimes k}$};

  \draw[line width=1.2pt] (-0.75,0) -- (box.west);
  \draw[line width=1.2pt] (box.east) -- (0.75,0);

  \node at (1,0) {$\sigma$};
  \node at (1.4,0) {$=$};
\node at (1.75,0) {$\pi$};

  \draw[black, line width=1.2pt] (2,0) -- (2.4,0);
  \draw[black, line width=1.2pt] (3.0,0) -- (3.4,0);
  \filldraw[gray!50, draw=black, line width=1.2pt] (2.7,0) circle (0.3);
  \node at (3.75,0) {$\sigma$};

  \node at (2.7,0) {$\tilde{\mathrm{G}}$};\label{fig:gram_doped}\;.
\end{tikzpicture}\;.
\end{equation}

\section{Results}
Our main interest lies in \emph{doped random Clifford circuits}, namely, quantum circuits composed of Clifford unitaries $C \in \mathcal{C}_N$ interspersed with a controllable number of non-Clifford gates $K$~\cite{Leone2021quantumchaosis}. 

We mainly focus on a circuit architecture that allows for analytical control: a sequence of $N_T + 1$ layers of global random Clifford operations on $\mathcal{H}$, interleaved with $N_T$ single-qudit non-Clifford gates $K$ applied on the first site\footnote{Since the permutation group on $N$ elements is a subgroup of the Clifford group, our results generalize trivially to the case where $K$ is applied to a random qudit.}. 
The resulting state $|\Psi(N_T)\rangle=C^{(N_T)}\prod_{k=0}^{N_T-1}  (K\otimes I^{\otimes N-1})C^{(k)}|0\rangle$ is graphically represented by 
\begin{equation}
\begin{tikzpicture}[scale=1, every node/.style={scale=1}]
\tikzset{block/.style={draw, very thick, fill=quantumpurple1, minimum width=1.cm, minimum height=2.5cm, rounded corners}}
\tikzset{dot/.style={circle, draw, fill=black, minimum size=6pt, inner sep=0pt}}
\tikzset{dot_brown/.style={circle, draw, fill=bluegray, minimum size=6pt, inner sep=0pt}}

\foreach \y in {0} {
  \draw[thick,black] (-0.5, \y) -- (9.5, \y);
}
\foreach \y in {-0.7} {
  \draw[very thick,black] (-0.5, \y) -- (9.5, \y);
}
\foreach \x in {1}{
\draw[block] (\x-1, 0.4) rectangle (\x, -1.2);
  \node[anchor=east] at (-0.5, -0.35){$|\Psi(N_T)\rangle\equiv $};
}

\foreach \x in {3,5,7,9}{
\draw[block] (\x-1, 0.4) rectangle (\x, -1.2);
}

\node[anchor=west] at (9.5, -0.35) {$|\Psi_0\rangle$};

\foreach \x in {1.5, 3.5, 5.5,7.5} {
  \foreach \y in {0} {
    \node[dot_brown,very thick] at (\x, \y) {};
  }
}

\draw[decorate, decoration={brace, amplitude=5pt}]
  (1.2, 0.6) -- (7.8, 0.6) node[midway, yshift=12pt] {$N_T$ times};

\end{tikzpicture}\;\label{eq:dopedstate}
\end{equation}
with the blue gate being $K$ and the purple independent, randomly distributed Clifford gates $C^{(k)}$ acting on $N$ qudits.
For this architecture, we denote by $N_T$ also the number of layers in the circuit and define the \emph{doping ratio} $q = N_T / N$ as the key parameter controlling the circuit's deviation from pure Clifford evolution. The initial state is chosen as $|\Psi_0\rangle = |0\rangle$, but by virtue of Clifford Haar invariance, our results generalize to any initial stabilizer state $|\Psi_0\rangle \in \mathrm{STAB}_{d,N}$.

\subsection{Magic spreading}\label{sec:magic_spreading}
Our first goal is to understand the propagation of magic resources in doped Clifford circuits.
For this scope, we consider the class of generalized stabilizer entropies (GSE), naturally arising from the algebraic structure of the Clifford group~\cite{turkeshi2025magicspreadingrandomquantum,magni2025anticoncentrationcliffordcircuitsbeyond}.
As discussed in Ref.~\cite{turkeshi2025magicspreadingrandomquantum}, these constitute a good measure of magic for many-body systems and include the stabilizer entropy as a particular example. 
The key observation lies in the distinction between the Haar and Clifford $k$-th commutants.
For $N \ge k - 1$, the Haar commutant has dimension $k!$, while the dimension of the Clifford commutant depends explicitly on the qudit dimension $d$, and is given by Eq.~\eqref{eq:comm_dim}.
Thus, while the two commutants coincide up to $k = 2$ for any $d$, already $k = 3$ (or $k = 4$) copies are sufficient to distinguish generic unitaries from Clifford unitaries in the case of qudits with $d \ge 3$ (or qubits, respectively).
This implies that any non-permutation, or intrinsic, commutant element $\Omega\in \overline{\mathrm{Comm}_k}(\mathcal{C}_N)\equiv \mathrm{Comm}_k(\mathcal{C}_N)\setminus \mathrm{Comm}_k(\mathrm{U}(D))$ can distinguish stabilizer states from non-stabilizer ones. 
In essence, such operators induce a magic measure, the so-called generalized stabilizer purity and entropy, which are respectively given by
\begin{equation}
    \zeta_\Omega(|\Psi\rangle) = \mathrm{Tr}\left( \Omega |\Psi\rangle\langle \Psi|^{\otimes k} \right)\;,\quad \mathcal{M}_\Omega(|\Psi\rangle) = -\log \zeta_\Omega(|\Psi\rangle)\;.
    \label{eq:GSEdef}
\end{equation}
The GSE $\mathcal{M}_\Omega(|\Psi\rangle)$ is (i) non-negative, and vanishes if and only if $|\Psi\rangle$ is a stabilizer state, (ii) is invariant by Clifford rotations $\mathcal{M}_\Omega(C|\Psi\rangle)=\mathcal{M}_\Omega(|\Psi\rangle)$, (iii) is additive $\mathcal{M}_\Omega(|\Psi\rangle\otimes |\Phi\rangle)=\mathcal{M}_\Omega(|\Psi\rangle)+\mathcal{M}_\Omega(|\Phi\rangle)$.
An important example is the operators 
\begin{equation}
    \Omega_{2\alpha}\equiv \frac{1}{D} \sum_{P\in \mathcal{P}_N}\left(P\otimes P^\dagger\right)^{\otimes \alpha}
\end{equation}
which introduce the stabilizer R\'enyi entropy~\cite{bittel2025completetheorycliffordcommutant,leone2022stabilizer}. 
We study the circuit average of the stabilizer purities
\begin{equation}
    \zeta_\Omega\equiv \mathbb{E}[\zeta_\Omega(|\Psi_{N_T}\rangle)]=\mathbb{E}[\llangle \Omega^\dagger|\rho^{\otimes k}_{N_T}\rrangle]=\llangle \Omega^\dagger|\mathcal{R}_k\rrangle\;,
\end{equation}
where we recasted Eq.~\eqref{eq:GSEdef} in terms of the replica tensor network formalism, and in the last step we defined the average $k$-replica state $|\mathcal{R}_k\rrangle=\mathbb{E}[|\rho^{\otimes k}_{N_T}\rrangle]$ with $\rho_{N_T}=|\Psi_{N_T}\rangle\langle \Psi_{N_T}|$ the replica state of the system. 
Using the graphical notation, we have 
\begin{equation}
\begin{tikzpicture}[scale=1, every node/.style={scale=1}]
\tikzset{block/.style={draw, very thick, fill=quantumpurple3, minimum width=1.cm, minimum height=2.5cm, rounded corners}}
\tikzset{dot/.style={circle, draw, fill=black, minimum size=6pt, inner sep=0pt}}
\tikzset{dot_brown/.style={circle, draw, fill=quantumpurple4, minimum size=6pt, inner sep=0pt}}
\tikzset{dot_brown2/.style={circle, draw, fill=gray!50, minimum size=6pt, inner sep=0pt}}

\foreach \y in {0} {
  \draw[thick,black] (-0.5, \y) -- (9.5, \y);
}
\foreach \y in {-0.7} {
  \draw[very thick,black] (-0.5, \y) -- (9.5, \y);
}
\foreach \x in {1}{
\draw[block] (\x-1, 0.4) rectangle (\x, -1.2);
  \node[anchor=east] at (-0.5, -0.35){$|\mathcal{R}_k\rrangle=$};
}

\foreach \x in {3,5,7,9}{
\draw[block] (\x-1, 0.4) rectangle (\x, -1.2);
}
\foreach \y in {0, -0.7} {
  \node[dot] at (9.5, \y) {};
  \node[anchor=west] at (9.8, \y) {$|0, 0\rrangle^{\otimes k}$};
}

\foreach \x in {1.5, 3.5, 5.5,7.5} {
  \foreach \y in {-0.7} {
    \node[dot_brown,very thick] at (\x, \y) {};
  }
}

\foreach \x in {1.5, 3.5, 5.5,7.5} {
  \foreach \y in {0} {
    \node[dot_brown2,very thick] at (\x, \y) {};
  }
}

\draw[decorate, decoration={brace, amplitude=5pt}]
  (1.2, 0.6) -- (7.8, 0.6) node[midway, yshift=12pt] {$N_T$ times};

\end{tikzpicture}\;
\label{eq:replicadopedstate}
\end{equation}
where we introduced the Gram matrices between commutant vector contractions, cf. Eq.~(\ref{fig:transfer_matrix}, \ref{fig:gram},\ref{fig:gram_doped}). 
As a result, the main object to compute is the replica operator 
\begin{equation}
    \mathcal{B}_{\tau,\sigma} = \sum_{\pi\in \Sigma_k(d)} \mathrm{Wg}_{\tau,\pi}(d^N) \mathrm{G}_{\pi,\sigma}(d^{N-1})\tilde{\mathrm{G}}_{\pi,\sigma}(d)\;,
    \label{eq:transfer_matrix}
\end{equation}
which is applied $N_T$ times on the system. 
Furthermore, the first contraction simplifies since $\bbra{\pi}(\kket{0,0}^{\otimes k}) = 1$ for all $\pi \in \Sigma_k(d)$, which allows us to use Eq.~\eqref{eq:weing_sum} to resum the first Weingarten and obtain the multiplicative factor $\mathcal{G}_{d,k,N}^{-1}$.
These remarks culminate in the final form
\begin{equation}
 \zeta_\Omega=\llangle \Omega|\mathcal{R}_k\rrangle=\mathcal{G}^{-1}_{d,k,N} \sum_{\pi,\sigma\in \Sigma_k(d)}\mathrm{G}_{\Omega,\pi}(d^N)\left(\mathcal{B}^{N_T} \right)_{\pi,\sigma},
 \label{eq:zeta_general}
\end{equation}
where, with a slight abuse of notation, we identify $\Omega$ with the associated stochastic Lagrangian subspace, and $\mathrm{G}_{\Omega,\pi}(d^N)$ comes from the last contraction, cf. Eq.~\eqref{eq:replicadopedstate}.
All the matrices in Eq.~\eqref{eq:transfer_matrix} are known once one has defined the local dimension, the number of replicas, and the doping gate $K$. In the following, we examine various cases by adjusting our defining parameters.

The previous discussion highlighted that three replicas ($k = 3$) are sufficient to reveal magic spreading in odd and prime qudit systems. Thus, we begin by specializing to this case and study the evolution of the generalized stabilizer purity associated with the corresponding intrinsic commutant operator. We then extend our analysis to broader setups and derive exact analytic expressions for generalized stabilizer purities at arbitrary replica number and qudit dimension $d\ge 3$.

\paragraph{Generic rotations.}
We begin by investigating the magic properties of doped Clifford circuits subjected to $Z$-rotations of the form $K = Z^\theta$, with $\theta \in [0,1)$. Our focus is on the three-replica commutant operator
\begin{equation}
\Omega_d \equiv \frac{1}{D} \sum_{P \in \mathcal{P}_N} P \otimes P \otimes P^\dagger\;,
\label{eq:Omega_d}
\end{equation}
for which we compute the corresponding generalized stabilizer purity at prime odd dimensions $d = 3, 5, 7$. For a detailed analysis of the structure of the Clifford commutant with three replicas, we refer to~\cite{zhu2024momentsquditcliffordorbits, nezami2020multipartite}. 

After straightforward but involved algebra, one finds that, in all considered cases, the generalized stabilizer purity exhibits the same functional form
\begin{equation}
\zeta_{\Omega_d}(\theta) = \zeta_{\Omega_d}^\mathrm{Haar} + \frac{d^N - 1}{d^N + 2} \left( \frac{\mu_d(\theta)\, d^{2N} - \beta_d(\theta)\, d^N - 1}{d^{2N} - 1} \right)^{N_T}\;,
\label{eq:gen_pur}
\end{equation}
where $\zeta_{\Omega_d}^\mathrm{Haar} = 3/(d^N + 2)$ is the expected value for Haar-random states~\cite{turkeshi2023measuring}, and the functions $\mu_d(\theta)$ and $\beta_d(\theta)$ capture the non-Clifford features introduced by the $Z^\theta$ rotation.

Their explicit forms are given by
\begin{equation}
\mu_d(\theta) = \frac{2}{d} + \xi_d(\theta), \qquad \beta_d(\theta) = (d - 2) \left( \frac{2}{d} - 2\xi_d(\theta) \right)\;,
\label{eq:mu_beta}
\end{equation}
with $\mu_d(\theta) = \zeta_{\Omega_d}(|\theta\rangle)$ interpreted as the generalized stabilizer purity of the single-qudit state $|\theta\rangle = Z^\theta |+\rangle$, where $|+\rangle = \sum_{i=1}^d |i\rangle/\sqrt{d}$.

The functions $\xi_d(\theta)$ are dimension-dependent and given by $\xi_3(\theta) = \left(1 + 2\cos(3\pi\theta)\right)^2$, $\xi_5(\theta) = \left(3 + 2\cos(2\pi\theta)\right)^2$ and $\xi_7(\theta) = (113 + 120\cos(2\pi\theta) + 12\cos(4\pi\theta))/{5}$.

\paragraph{Qudit $T$-gates.}\label{sec:T-gate}
A similar analysis applies to the generalization of $T$-gates to qudit systems. Following~\cite{Prakash_2018, howard2012qudit}, these second-level Clifford hierarchy gates are denoted by $T_3 \equiv Z^{1/3}$ and $T_d \equiv \sum_k \omega^{k^3 6^{-1}} \ket{k}\bra{k}$ for $d\geq 3$, with the inverse done mod$\,d$. As previously discussed, such operators can be injected into the system in a fault-tolerant manner via the teleportation protocol~\cite{gottesmann1999demonstrating}. 
Let us first specialize to three replicas. By direct inspection we find that, for any $\Omega \in \overline{\mathrm{Comm}_3}(\mathcal{C}_N)$ and for $d = 3$ and $d \equiv 2 \mod 3$, Eq.~\eqref{eq:gen_pur} remains valid provided one sets $\xi_d(\theta) = 0$. 

While the above analysis systematically characterizes magic spreading for $k = 3$ replicas, we have also performed exact computations of the Clifford commutant for the case $d = 3$ and $k = 4$. Once again, Eq.~\eqref{eq:gen_pur} applies to the two non-trivial commutant elements
\begin{equation}
\Omega_{3,1} = \frac{1}{D} \sum_P P^{\otimes 3} \otimes I\;, \qquad \Omega_4 = \frac{1}{D} \sum_P (P \otimes P^\dagger)^{\otimes 2}\;,
\label{eq:omega_4}
\end{equation}
where $\Omega_{3,1}$ corresponds to the embedding of the $k = 3$ commutant element $\Omega_3$, and $\Omega_4$ is the standard stabilizer purity operator. The main difference lies in the values of $\xi_d$: we find $\xi_d = 0$ for $\Omega_{3,1}$ and $\xi_d = -1/9$ for $\Omega_4$.

\paragraph{General formula for magic spreading.}

The exact results presented above suggest a simplified and universal expression for magic spreading in doped Clifford circuits, regardless of the choice of GSE.
Indeed, in the large-$N$ limit, we find
\begin{equation}
    \zeta_{\Omega} \simeq \zeta_{\Omega}^\mathrm{Haar} + \left( \zeta_\Omega(|K\rangle) \right)^{N_T}\;,
    \label{eq:gen_pur_scaling}
\end{equation}
where $\zeta_\Omega(|K\rangle)=\mathrm{tr}[\Omega K^{\otimes k} \Omega (K^\dagger)^{\otimes k}]$ quantifies the magic of the single-qudit state $|K\rangle$. We now show that this scaling form holds for arbitrary local dimension $d$ and replica number $k$.

In the thermodynamic limit, vectors in the Clifford commutant become approximately orthogonal, as $\mathrm{G}_{\pi,\sigma}(D) \simeq D^k\, \delta_{\pi,\sigma} + \mathcal{O}(D^{k-1})$~\cite{bittel2025completetheorycliffordcommutant}. 
Consequently, to leading order in $N \gg 1$, the transfer matrix simplifies to
\begin{equation}
    \mathcal{B}_{\pi,\sigma} = \delta_{\pi,\sigma} \times \begin{cases}
        1 & \sigma \in \mathrm{S}_k\;, \\
        \zeta_\Omega(|K\rangle) & \sigma \in \Sigma_k(d)\setminus \mathrm{S}_k\;.
    \end{cases}
\end{equation}
Plugging this into Eq.~\eqref{eq:zeta_general}, we obtain
\begin{equation}
    \zeta_\Omega \simeq \frac{1}{d^{kN}} \sum_{\pi,\sigma \in \Sigma_k(d)} \mathrm{G}_{\Omega,\pi}(d^N)\, \delta_{\pi,\sigma} \left[ \Theta(\sigma \in \mathrm{S}_k) + \zeta_\Omega(|K\rangle)^{N_T} \, \Theta(\sigma \notin \mathrm{S}_k) \right]\;,
\end{equation}
where $\Theta(x \in A)$ is the indicator function, equal to 1 if $x \in A$ and 0 otherwise. 
Simplifying the sum yields
\begin{equation}
    \zeta_\Omega \simeq \frac{1}{d^{kN}} \left(\sum_{\pi \in \mathrm{S}_k} \mathrm{G}_{\Omega,\pi}(d^N)\right) + \zeta_\Omega(|K\rangle)^{N_T}\;,
\end{equation}
where the first term corresponds to the leading-order contribution of stabilizer purity for Haar-random states, as detailed in App.~\ref{AppendixA}. 
Thus, we arrive at a general scaling formula for any generalized stabilizer purity Eq.~\eqref{eq:gen_pur_scaling}, from which the magic spreading of doped Clifford circuits follows as
\begin{equation}
    \mathcal{M}_\Omega \simeq -\log\left[ \zeta_\Omega^\mathrm{Haar} + \zeta_\Omega(|K\rangle)^{N_T} \right]\;.
    \label{eq:gse_scaling}
\end{equation}
As a further benchmark, we note this formula correctly reproduces previously known results for $d = 2$ and proves a recent conjecture regarding stabilizer entropy proposed in~\cite{haug2024probingquantumcomplexityuniversal}.

\begin{figure}
    \centering
    \includegraphics[width=0.9\linewidth]{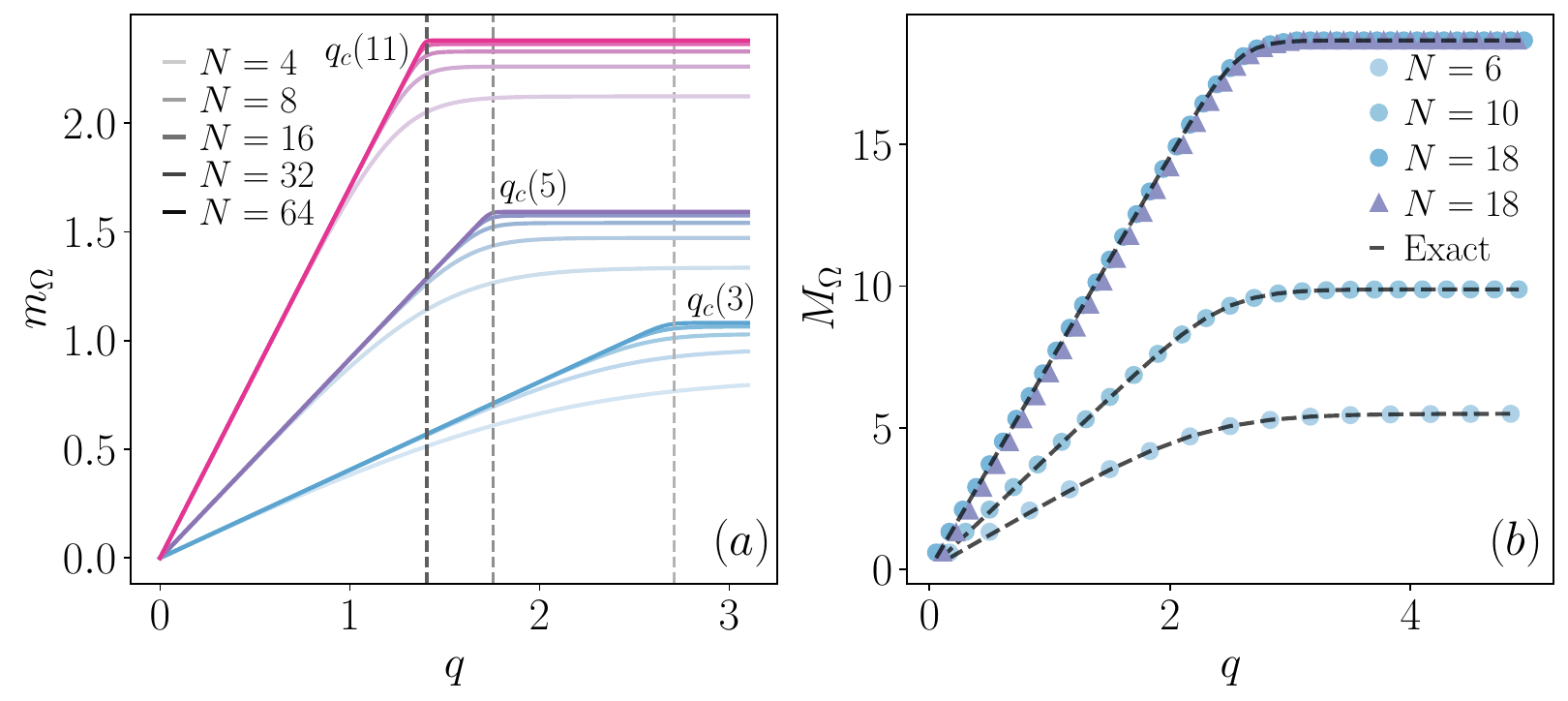}
    \caption{$(a)$ GSE densities for different local dimensions (d=3,5,11) and system sizes computed with the reference $T_d$-gate. The initial linear growth culminates in a transition, at the value $q_c$ in Eq.~\eqref{eq:critical_q}, that becomes sharper as the system size increases. After this point, the universal value $\log(d)$ is reached for $N\rightarrow \infty$. $(b)$ We observe the exact superposition of the numerical value of the GSE at $d=3$ and $k=3$ obtained from the local doped circuit (blue dots) and local doped staircase circuit (purple triangles) with the exact values given by Eq.~\eqref{eq:gen_pur} with $\xi_d(\theta)=0$. Due to the different construction and layer definition, the density $q$ of the staircase circuit has been doubled to match, per layer, the one of the brickwork. This result highlights the marginal role of locality for the spreading of magic in doped Clifford circuits.}
    \label{fig:doped_magic}
\end{figure}

\paragraph{Dynamical phase transition.} 
The main result, Eq.~\eqref{eq:gen_pur}, highlights a dynamical phase transition for the spreading of magic in generic doped Clifford circuits and for any generalized stabilizer entropy. 
Specifically, fixing the doping rate $q=N_T/N$, the magic density $m_\Omega = \mathcal{M}_\Omega/N$ increases linearly with the doping ratio for various local dimensions and system sizes, until it reaches a critical value $q_c$ (see Fig.~\ref{fig:doped_magic}$(a)$). 
Beyond this point, the magic density saturates to a universal value given by $m_{\Omega,\mathrm{max}} \approx \log(d)$ in the scaling limit.

Interestingly, the critical doping threshold $q_c$ decreases with increasing local dimension. Indeed, taking the derivative of Eq.~\eqref{eq:gse_scaling}, one finds
\begin{equation}
    q_c = -\frac{\log d}{\log \zeta_\Omega(|K\rangle)}\;,
    \label{eq:critical_q}
\end{equation}
which implies that higher-dimensional qudits lead to faster growth of circuit complexity and earlier onset of chaotic behavior.
In Fig.~\ref{fig:doped_magic}(a) we present how the magic density varies in terms of the doping ratio for the reference qudit $T_d$-gate, cf. Sec.~\ref{sec:T-gate}, considering different local dimensions. We can clearly observe the shift of the threshold to smaller values of $q$ as $d$ increases and how the transition becomes sharper, tending to the Haar value, with $N\rightarrow \infty$.

\paragraph{Numerical checks with local circuits.}

We now test the validity of our analytical formula against local random circuits with a brickwork and a staircase architecture. In the first instance, we consider a system of $L = 2n + 2$ sites, and for concreteness, we focus on $k = 3$ replicas and local dimension $d = 3$. The circuit consists of alternating layers of randomly chosen two-qudit Clifford gates acting on neighboring sites, forming a standard brickwork pattern. This setting contrasts with the global Clifford circuits discussed previously, and allows us to probe the role of locality in the spreading of magic. 

To introduce non-Clifford resources, every second layer features a doped gate inserted at a fixed central position. Specifically, we replace the central two-qudit gate with the product $(T_{3,L/2} \otimes T_{3,L/2+1}) C_{L/2,L/2+1}$, where $T_{3,i}$ are the reference $T_3$-gate acting on site $i$, and $C_{L/2,L/2+1}$ is a two-body Clifford gate. 
This doping is applied periodically and consistently throughout the circuit, ensuring a uniform injection of non-Cliffordity—equivalent to one doped gate per layer on average. The resulting architecture is

\begin{equation}\label{eq:rmpsBrickwallRotatedRight13GreyMiddle}
\begin{tikzpicture}[baseline=(current bounding box.center),scale=0.6]

\def\nqubits{6}
\def\ndepth{13}

\foreach \y in {1,...,\nqubits}{
    \draw[thick] (0,\y) -- (\ndepth+2.,\y);
    \draw[thick, fill=white] (\ndepth+2.,\y) circle (0.2);
    \node[scale=0.75] at (\ndepth+2.75,\y) {\large $|0\rangle$};
}

\foreach \d in {0,...,\numexpr\ndepth-1} {
    \pgfmathsetmacro{\x}{\d + 1} 
    \ifodd\d
        \foreach \y in {1,3,5} {
            \ifnum\y=3
                \def\gatecolor{bluegray}
            \else
                \def\gatecolor{quantumpurple1}
            \fi
            \draw[thick, fill=\gatecolor, rounded corners=2pt]
                (\x-0.3,\y-0.2) rectangle (\x+0.3,\y+1.2);
        }
    \else
        \foreach \y in {2,4} {
            \draw[thick, fill=quantumpurple1, rounded corners=2pt]
                (\x-0.3,\y-0.2) rectangle (\x+0.3,\y+1.2);
        }
    \fi
}

\foreach \d in {\numexpr\ndepth} {
    \pgfmathsetmacro{\x}{\d + 1} 
    \ifodd\d
        \foreach \y in {1,3,5} {
            \ifnum\y=3
                \def\gatecolor{quantumpurple1}
            \else
                \def\gatecolor{quantumpurple1}
            \fi
            \draw[thick, fill=\gatecolor, rounded corners=2pt]
                (\x-0.3,\y-0.2) rectangle (\x+0.3,\y+1.2);
        }
    \else
        \foreach \y in {2,4} {
            \draw[thick, fill=quantumpurple1, rounded corners=2pt]
                (\x-0.3,\y-0.2) rectangle (\x+0.3,\y+1.2);
        }
    \fi
}

\end{tikzpicture}\;.
\end{equation}

To analyze the resulting dynamics, we employ the replica tensor network framework introduced in~\cite{magni2025anticoncentrationcliffordcircuitsbeyond}, which allows us to track the evolution of the generalized stabilizer purity associated with the commutant operator $\Omega$. We translate the circuit into the replica space and compute the magic spreading as a function of the number of doped layers~\cite{turkeshi2025magicspreadingrandomquantum}.

As an additional verification, we analyze a different local architecture: a staircase circuit with local dimension $d=3$ and $k=3$ replicas. The system consists of $L$ sites. 
For simplicity, we first apply a single layer of two-qutrit random Clifford gates across the chain. 
The subsequent dynamics proceed in a staircase pattern, where each timestep $t$ is defined by a sequence of $L-1$ two-qutrit random Clifford gates, $C_\mathrm{step}(\tau)=(\prod_{i=1}^{\tau-1} C_{i,i+1})[(T_{3,\tau} \otimes T_{3,\tau+1}) C_{\tau,\tau+1}](\prod_{j=\tau+1}^{L-1} C_{j,j+1})$, where we substituted at position $0\le \tau \le L-1$ a magic gate. 
Including the boundary conditions, the final dynamics is generated by $C_t = \prod_{s=1}^t C_\mathrm{step}(s \bmod (L-1))$.

Remarkably, we find that the numerical results for these local circuits match the analytical predictions derived for global Clifford dynamics with high precision, see Fig.~\ref{fig:doped_magic}(b). This agreement is nontrivial: although the circuit is now composed of spatially local gates, and the doping occurs at a fixed location rather than globally, the overall magic dynamics remains unchanged. In particular, the scaling of the magic is entirely captured by our formula, with the same exponential growth in the number of doped layers.

This observation highlights an important distinction: for state magic, the spatial location of the non-Clifford resource is irrelevant--as long as the doping occurs periodically, the system's complexity grows identically to the global case. 
We note that this stands in contrast with operator magic, where the locality of the gate action is crucial for determining the spreading behavior~\cite{dowling2024magic}. 
Our results, therefore, not only validate the analytic formula but also underscore its robustness across different circuit architectures.

\subsection{Anticoncentration and entanglement features}
We now complement our analysis of the complexity in doped Clifford circuits from the perspective of anticoncentration and entanglement. 
We begin by presenting exact results for the anticoncentration, quantified by the ensemble-averaged inverse participation ratios and the participation entropies. Focusing on qutrit systems, we establish a connection between these findings and the resource theory of magic via the multifractal (anti)flatness -- an indicator of magic that builds on the notion of Clifford orbits and the fluctuations of the inverse participation entropy. 
Subsequently, we apply an analogous study to the $k$-Rényi purities and entanglement entropies, connecting them to the generalized stabilizer entropies through the concept of entanglement antiflatness, for $d=3$ systems.

\paragraph{Anticoncentration of coherences.}
Generic unitaries map a localized state, expressed in a chosen reference basis, into a superposition of many basis states. 
As a result, the probability distribution becomes broadly spread in this basis, and the circuit ensemble is said to be anticoncentrated if its output statistics resemble those of a stationary random ensemble. For chaotic systems, this asymptote is the Porter-Thomas distribution~\cite{porter1956fluctuations}.
Anticoncentration is closely related, but not identical, to the concept of Fock space or Hilbert space delocalization. The latter refers to the degree to which a single pure state $|\Psi\rangle$ is spread across the computational basis $\{ |x\rangle \}$ of the Hilbert space. This property is quantified by the inverse participation ratios (IPRs) and participation entropy, respectively~\cite{luitz2014participation,mace2019multifractal,luitz2014universal,sierant2022universal,tirrito2024anticoncentrationmagicspreadingergodic}
\begin{equation}
    I_k(|\Psi\rangle) = \sum_{x=0}^{D-1} |\langle x|\Psi\rangle|^{2k },\quad H_k(|\Psi\rangle)=\frac{1}{1-k}\log\left[I_k(|\Psi\rangle)\right]\;,
\end{equation}
where $k \geq 2$ and $D$ is the total Hilbert space dimension. 
Large values of $H_k(|\Psi\rangle)$ indicate a high degree of delocalization, i.e., that the state has support over many basis elements.
In contrast, the notion of anticoncentration refers to an ensemble of states, typically generated by random or pseudorandom quantum circuits~\cite{dalzell2022random,schuster2025randomunitariesextremelylow,fefferman2024anticoncentrationunitaryhaarmeasure}. Accordingly, one defines the ensemble-averaged IPRs as \footnote{Here and throughout, we use the same notation for ensemble-averaged quantities, omitting explicit reference to the state.}
\begin{equation}
    I_k=\mathbb{E}_{\Psi}\left[\sum_{x=0}^{D-1} |\langle x|\Psi\rangle|^{2k }\right]\;.
\end{equation}
Notably, anticoncentration plays a fundamental role in benchmarking computational quantum advantage via cross-entropy benchmarks~\cite{morvan2024phase,arute2019quantum,boixo2018characterizing}. 
For Clifford circuits with initial magic states, Ref.~\cite{magni2025anticoncentrationcliffordcircuitsbeyond} presented a complete solution for random tensor networks and shallow circuits. 

We now extend this analysis to global Clifford interspersed with magic gates. First, we note that
\begin{equation}
    I_k=\sum_{x=0}^{D-1} \left(\llangle x,x|^{\otimes k}\right) |\mathcal{R}_k\rrangle= D\left(\llangle 0,0|^{\otimes k}\right) |\mathcal{R}_k\rrangle\;,
\end{equation}
where in the first step we used simple algebraic manipulation and the definition of $|\mathcal{R}_k\rrangle$ in Eq.~\eqref{eq:replicadopedstate}, and in the second step we used Clifford invariance to reabsorb the bitflips determining $x$. 
Recalling the transfer matrix definition Eq.~\eqref{eq:transfer_matrix}, we obtain
\begin{equation}
    I_k=D \;\mathcal{G}^{-1}_{k,d,N}\;\sum_{\pi,\sigma\in \Sigma_k(d)} \mathcal{B}_{\pi,\sigma}^{N_T}\;,
\end{equation}
which can be resolved exactly once the properties of $\mathcal{B}$ are known. 

Let us now focus on $k=3$, albeit similar results extend to higher replicas. 
For generic rotations $K=Z^\theta$, the exact results for $d=3,5,7$ is given by 
\begin{equation}
    I_3(\theta)=I_3^\mathrm{Haar}\left[1+\frac{(d-2)(d^N-1)}{3(d^N+d)} \left(\frac{\mu(\theta)d^{2N}-\beta(\theta)d^N-1}{d^{2N}-1}\right)^{N_T}\right]\;,
\end{equation}
where $I_3^\mathrm{Haar}=6/[(D+1)(D+2)]$, and $\mu_d(\theta)$ and $\beta_d(\theta)$ are defined as in Sec.~\ref{sec:magic_spreading}, cf. Eq.~\eqref{eq:mu_beta}. 
Similarly, for the reference qudit $T_d$-gate, we obtain the same functional form when $\xi_d=0$. 

These results demonstrate that the relative error between the typical value $I_3(|\Psi(N_T)\rangle)$ and the Porter-Thomas benchmark $I_3^\mathrm{Haar}$ decays exponentially. At leading order, and for generic $k$ and odd prime $d \ge 3$, the quasi-orthogonality condition yields
\begin{equation}
I_k \simeq I_k^\mathrm{Haar} \left(1 + \frac{\sum_{\Omega \in \overline{\mathrm{Comm}_k}(\mathcal{C}_N)} [\zeta_{\Omega}(|K\rangle)]^{N_T}}{k!} \right)\;,
\end{equation}
which mirrors the form derived in Ref.~\cite{magni2025anticoncentrationcliffordcircuitsbeyond}. In particular, a doped circuit with $N_T \sim \log(N)$ non-Clifford gates suffices to achieve Haar-like values within relative error $\varepsilon$, consistent with the findings in~\cite{magni2025anticoncentrationcliffordcircuitsbeyond}.

\paragraph{Multifractal (Anti)flatness.}
We now consider the multifractal flatness, an antiflatness measure introduced in Ref.~\cite{turkeshi2023measuring}, which is proportional to the linearized stabilizer entropy for qubit systems.

This quantity is defined for a state $|\Psi\rangle$ through averaging over its Clifford orbit
\begin{equation}
    \mathcal{F}(|\Psi\rangle)\equiv \mathbb{E}_{C}[I_3(C|\Psi\rangle)- I_2^2(C|\Psi\rangle)]\;.
\end{equation}
We focus on the case $d = 3$ (qutrits), though the generalization to higher-dimensional qudits is straightforward. 
For qutrits, the Clifford average is given by
\begin{equation}
    \mathbb{E}_C[I_3(C|\Psi\rangle)] = \frac{6+ 2\zeta_{\Omega_3}}{(d^N+1)(d^N+d)},
\end{equation}
where $\zeta_{\Omega_3} = \llangle \Omega_3 | (|\Psi,\Psi\rrangle^{\otimes 3})$ is the generalized stabilizer purity associated with the commutant operator defined in Eq.~\eqref{eq:Omega_d}. 
For the squared average, we use $\Omega_4$ from Eq.~\eqref{eq:omega_4}, yielding the explicit expression 
\begin{equation}
    \mathbb{E}_C[I_2^2(C|\Psi\rangle)] = \frac{4(1+ \zeta_{\Omega_4})}{(d^N+1)(d^N+d)}\;.
\end{equation}
Combining the two contributions, we find
\begin{equation}
    \mathcal{F}(|\Psi\rangle)=\frac{2}{(d^N+1)(d^N+d)}\left[1+\zeta_{\Omega_3}(|\Psi\rangle) - 2 \zeta_{\Omega_4}(|\Psi\rangle)\right]\;.\label{eq:antitt}
\end{equation}
Notably, this quantity is not merely proportional to the stabilizer Rényi entropy, but involves a nontrivial combination of multiple commutant operators. These results also clarify that the protocol proposed in Ref.~\cite{haug2024efficient} does not directly extend to the qutrit case considered here. Specifically, the term $I_3$ depends on the observable $\Omega_3$, which is proportional to a projector in replica space. This structure leads to a non-flat spectrum, complicating efficient estimation and requiring access to the complex-conjugated state. By contrast, the contribution $I_2^2$ involves the operator $\Omega_4$, which admits a more tractable form and allows for straightforward evaluation within our framework.

\paragraph{Entanglement.}
Next, we study the bipartite entanglement of doped Clifford circuits. For this scope, we introduce a bipartition of a given state $|\Psi\rangle$ in two parts $X\cup X_c$, with $X_c$ the complement of $X$. 
The $k$-th purity of the reduced density matrix $\rho_X=\mathrm{tr}_{X_c}|\Psi\rangle\langle \Psi|$ and the associated entanglement entropies are given, respectively, by
\begin{equation}
    \mathcal{P}_k(|\Psi\rangle) = \mathrm{tr}(\rho_X^k),\quad \mathcal{S}_k = \frac{1}{1-k}\log\left[\mathcal{P}_k(|\Psi\rangle)\right]\;.
\end{equation}
The entanglement entropy measures the number of Bell pairs shared between $X$ and $X_c$~\cite{horodecki2009,amico,Calabrese_2009}. 
Let us now focus on the average entanglement purities of the doped Clifford circuit. Again, we rephrase the problem to a replica expectation value 
\begin{equation}
    \mathcal{P}_k = \mathbb{E}[\mathcal{P}_k(|\Psi_{N_T}\rangle)]=\llangle \mathrm{cyc}_k|\mathcal{R}_k\rrangle\;,
\end{equation}
where $\mathrm{cyc}_k$ is the cyclic operator on $X$, explicitly for any computational basis vector $|x\rangle=|x_X,x_{X_c}\rangle$ we have
\begin{equation}
    \mathrm{cyc}_k=\sum_{x^1,\dots,x^k=0}^{D-1} \bigotimes_{j=1}^k|x^j_X,x^j_{X_c}\rangle\langle x_{X}^{j+1},x_{X_c}^j|
\end{equation}
with $x^{k+1}\equiv x^1$. Up to reshaping, this operator can be expressed from $I^{\otimes N_X} \otimes T_{(12\dots k)}^{\otimes (N-N_X)}$, with $T_{(12\dots k)}$ the cyclic permutation on a single qudit. 
We can again replace the value of $|\mathcal{R}_k\rrangle$ in terms of the transfer matrices to obtain the final expression 
\begin{equation}
    \mathcal{P}_k =\mathcal{G}_{k,d,N}^{-1}\sum_{\pi,\sigma\in \Sigma_k(d)} \mathrm{G}_{\iota,\pi}(d^{N_X})\mathrm{G}_{\mathrm{cyc},\pi}(d^{N-N_X}) \mathcal{B}_{\pi,\sigma}^{N_T} \;,
\end{equation}
where $\iota$ is the index of the identity element in $\Sigma_k(d)$ and $\mathrm{cyc}$ that corresponding to the cyclic permutation $(12\dots k)\in \mathrm{S}_k\subset \Sigma_k(d)$. 

As for the anticoncentration features, we consider generic rotations $Z^\theta$, and study $d=3,5,7$ for $k=3$. 
A simple but involved computation leads to 
\begin{equation}
\mathcal{P}_3(\theta)=\mathcal{P}_3^\mathrm{Haar}\left(1+\frac{(d-2)(d^{2N_X}-1)(d^{2(N-N_X)}-1)}{(d^{2N_X}+ 3 d^{N}+d^{2(N-N_X)})(d^N+d)}\left(\frac{\mu_d(\theta)d^{2N}-\beta_d(\theta)d^N-1}{d^{2N}-1}\right)^{N_T}\right)\;,
\label{eq:purity}
\end{equation}
where again $\mu_d(\theta)$ and $\beta_d(\theta)$ are given by Eq.~\eqref{eq:mu_beta}, and 
\[
\mathcal{P}_3^\mathrm{Haar}\equiv \frac{d^{2N_X}+ 3 d^{N}+d^{2(N-N_X)}}{(d^N+1)(d^N+2)}\;,
\]
is the third purity of a random Haar state. 
In particular, similarly to the previous results for anticoncentration, a doped circuit with $N_T \sim \log(N)$ non-Clifford gates suffices to reproduce the Haar value of the third purity within relative error $\varepsilon$, exhibiting again pseudo-magic behavior. This observation is consistent with the fact that a global Clifford state is already maximally entangled, and only the subleading corrections depend on the non-Clifford doping.

\paragraph{Antiflatness of entanglement.}
Finally, in full analogy with the inverse participation ratio, we compute the entanglement antiflatness, defined as~\cite{tirrito2024quantifyingnon,rattacaso2023stab,cusumano2025nonstabilizernessviolationschshinequalities,viscardi2025interplayentanglementstructuresstabilizer,jasser2025stabilizerentropyentanglementcomplexity,iannotti2025entanglementstabilizerentropiesrandom}
\begin{equation}
    \mathcal{\mathcal{A}}(\ket{\Psi}) =\mathrm{Tr}(\rho_X^3)-  \mathrm{Tr}^2(\rho_X^2).
\end{equation}
As before, we evaluate the individual terms for $\rho_X(C)\equiv \mathrm{tr}_{X_c}[C|\Psi\rangle\langle \Psi|C^\dagger]$, yielding for $d=3$
\begin{equation}
    \mathcal{A}= \frac{(d^{2N_X}-1)(d^{2(N-N_X)}-1)}{(d^N+d)(d^{2N}-1)}\left[1+\zeta_{\Omega_3}(|\Psi\rangle) - 2 \zeta_{\Omega_4}(|\Psi\rangle)\right]\;.\label{eq:antif}
\end{equation}
As in the qubit case~\cite{tirrito2024quantifyingnon, turkeshi2023measuring}, both antiflatness measures, Eqs.~\eqref{eq:antif} and~\eqref{eq:antitt}, depend on the same functional combination of generalized stabilizer purities, differing only by a prefactor determined by the observable under consideration.
Remarkably, for both the entanglement antiflatness and the multifractal flatness, this prefactor coincides with the qubit result even in the qutrit case, suggesting that the functional form extends to higher-dimensional qudits—provided the generalized stabilizer purities are defined appropriately. We have verified this conjecture for d = 3, 5, and 11, and leave a comprehensive analysis to future work.

\subsection{Quantum chaos}
Moving from state properties to circuit-level dynamics, a central hallmark of quantum chaos is the breakdown of operator locality under unitary evolution.  
This phenomenon is captured by \emph{out-of-time-order correlators} (OTOCs)~\cite{Roberts_2017}, which serve as powerful diagnostics of scrambling, operator spreading, and complexity growth. 
Given a set of $k$ operator pairs $A_1,\dots,A_k$ and $B_1,\dots,B_k$, the associated $2k$-point OTOC is defined as
\begin{equation}
    \mathrm{OTOC}_{2k}(U) = \frac{1}{D}\mathrm{tr} \left( A_1 U B_1 U^\dagger \cdots A_k U B_k U^\dagger \right) \;,
    \label{eq:otoc2n}
\end{equation}
where $U$ is the unitary evolution and $D$ is the Hilbert space dimension. The decay of this correlator reflects the growth of non-commutativity under Heisenberg evolution and the spreading of initially local operators across the system.

In doped Clifford circuits, the behavior of $\mathrm{OTOC}_{2k}$ depends sensitively on the replica index $k$. Due to the design properties of the Clifford group, Clifford dynamics is indistinguishable from Haar-random evolution for small  $k$: the Clifford group forms a unitary 2-design for all local dimensions $d$, and a 3-design only for qubits ($d = 2$).  
As a result, for qudits with $d \geq 3$, Clifford circuits replicate Haar statistics only up to $k = 2$.  
Beyond this threshold—i.e., $k = 3$ for qudits or $k = 4$ for qubits—deviations from Haar behavior become detectable, and OTOCs provide a direct probe of these differences~\cite{zhu2016cliffordgroupfailsgracefully}.

Recent work~\cite{Leone2021quantumchaosis} has rigorously established that achieving Haar-like behavior in OTOCs—thereby signaling the onset of genuine quantum chaos—requires an extensive number $\Theta(N)$ of non-Clifford gates.  
Below this threshold, the circuit remains in a non-chaotic phase that is classically simulable and fails to approximate higher-order unitary designs. 
Here, we explore the onset of quantum chaos in qudit doped Clifford circuits, focusing on the minimal nontrivial case of $ k = 3 $ replicas.  
This corresponds to the six-point correlator $ \mathrm{OTOC}_6 $, which is already sensitive to non-Clifford effects for local dimensions $ d \geq 3 $.  
Following Ref.~\cite{Roberts_2017}, we consider the operator assignments $ A_1 = A_2 = P $, $ A_3 = (P^2)^\dagger $, and $ B_1 = B_2 = Q $, $ B_3 = (Q^2)^\dagger $, where $ P, Q $ are generalized Pauli operators, such that $ A_1 A_2 A_3 = B_1 B_2 B_3 = \mathbb{I} $.  
This choice yields an analytically tractable setup that nevertheless captures the essential features of the chaos transition and the growth of unitary design complexity.  
In particular, we show that the structure of the Clifford commutant at $ k = 3 $ enables an exact treatment via replica methods, and that the decay of $ \mathrm{OTOC}_6 $ under non-Clifford doping reveals the crossover from non-chaotic to chaotic dynamics.

\begin{figure}
    \centering
    \includegraphics[width=0.9\linewidth]{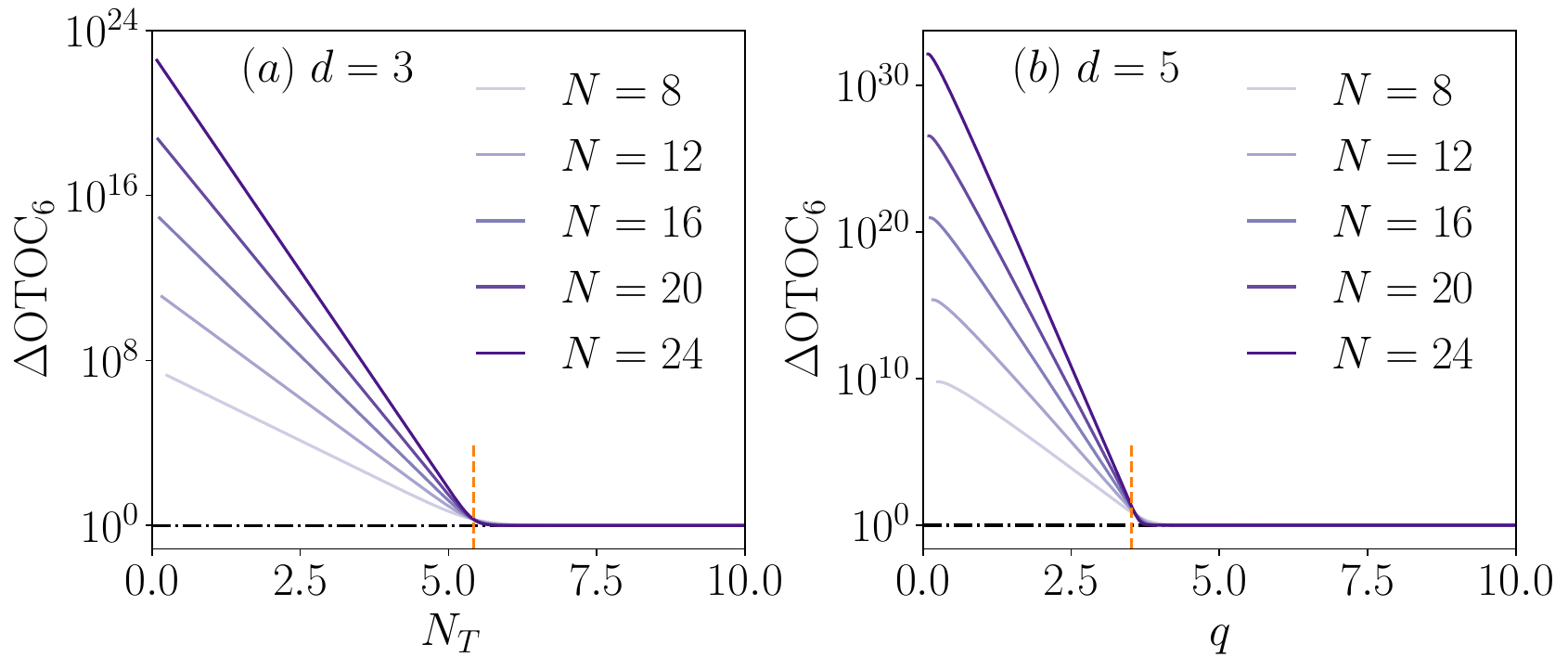}
    \caption{We illustrate the scaling of Eq.~\eqref{eq:DeltaOTOC} with the number of injected non-Clifford gates $ N_T $ for local dimensions (a) $ d = 3 $ and (b) $ d = 5 $. 
    Across different system sizes, we observe an exponential decay of $ \Delta \mathrm{OTOC}_6 $, which approaches the Haar-random value $ \mathrm{OTOC}_6^{\mathrm{Haar}} $ at a characteristic doping rate $q=N_T/N\sim O(1)$.  
    Evaluating the critical density marking the transition from the non-chaotic to the chaotic regime (dashed lines in orange), we find that it coincides with $ q_c^\mathrm{OTOC}=2q_c $, the critical doping density for the magic spreading, cf. Eq.~\eqref{eq:otocs}. After this point, the system reaches the Haar value ($\Delta \mathrm{OTOC}_6\mapsto 1$).
    This observation supports our previous findings on the accelerated growth of complexity at higher local dimensions. 
    }
    \label{fig:Dotoc6}
\end{figure}

Inserting the doped Clifford circuit of Eq.~\eqref{eq:dopedstate} as the unitary of interest and averaging over the circuit ensemble, we obtain the averaged OTOC
\begin{equation}
    \mathrm{OTOC}_6 = \frac{1}{d^N} \llangle T_{(123)}\, P \otimes P \otimes (P^2)^\dagger \,\Big|\, \mathbb{E}\left[(U_{N_T} \otimes U_{N_T}^*)^{\otimes 3}\right] \,\Big|\, Q \otimes Q \otimes (Q^2)^\dagger \rrangle\;,
\end{equation}
where the expectation is taken over all circuit realizations, and $ T_{(123)} $ denotes the cyclic permutation operator acting on the three replicas.  
As in the case of magic spreading, cf. Sec.~\ref{sec:magic_spreading}, we express this average using the replica transfer matrix formalism.  
A straightforward algebraic manipulation then yields
\begin{equation}
    \mathrm{OTOC}_6 = \frac{1}{d^N} \sum_{\tau,\pi,\sigma \in \Sigma_3(d)} \llangle T_{(123)}\, P \otimes P \otimes (P^2)^\dagger \,\big|\, \tau \rrangle \; \mathcal{B}_{\tau,\pi}^{N_T} \; \mathrm{Wg}_{\pi,\sigma}(d^N) \; \llangle \sigma \,\big|\, Q \otimes Q \otimes (Q^2)^\dagger \rrangle\;.
    \label{eq:otoc6_general}
\end{equation}

To proceed analytically, we specialize to $ d = 3 $ and choose as reference the $ T_3 $-gate, characterized by magic $ \mu = 2/3 $ and $ \xi_d = 0 $, cf. Sec.~\ref{sec:magic_spreading}.  
While our method extends to other local dimensions and gate choices, this setup already captures the essential features of the crossover. 
Carrying out the sum explicitly, we find
\begin{equation}
\begin{split}
    \mathrm{OTOC}_6 &= \mathrm{OTOC}_6^{\mathrm{Haar}} + \frac{d^{2N}(d^N+d)(d^N-2)}{2(d^{2N}-1)(d^{2N}-4)}\left(\frac{\mu\, d^{2N} - \mu\, d^N - 1}{d^{2N}-1}\right)^{N_T} \\
    &\quad + \frac{d^{2N}(d^N-d)(d^N+2)}{2(d^{2N}-1)(d^{2N}-4)}\left(\frac{\mu\, d^{2N} + \mu\, d^N - 1}{d^{2N}-1}\right)^{N_T}\;,
    \label{eq:otoc6d3}
\end{split}
\end{equation}
where we defined the Haar-averaged value of the OTOC
\begin{equation}
    \mathrm{OTOC}_6^{\mathrm{Haar}}=\frac{d^{2N}+4}{(d^{2N}-4)(d^{2N}-1)}\;,
    \label{eq:otocHaar}
\end{equation}
see App.~\ref{AppendixA} for details on this computation.  
In the large-$ N $ limit, the expression simplifies to
\begin{equation}
    \mathrm{OTOC}_6 \simeq \mathrm{OTOC}_6^{\mathrm{Haar}} + \mu^{N_T},
\end{equation}
exhibiting a simple exponential decay toward the Haar value as the number of doped layers $ N_T $ increases. 
For a sublinear number of doped gates, the OTOC remains pinned to its Clifford value.  
In contrast, once the number of doped layers becomes extensive, $ N_T = \Theta(N) $, the OTOC converges to its Haar-random value.  
This transition mirrors the rigorous findings of Ref.~\cite{Leone2021quantumchaosis}, where it was shown that $ \Theta(N) $ non-Clifford insertions are both necessary and sufficient to reproduce Haar-like behavior in higher OTOCs and in purity fluctuations.  
Notably, direct inspection reveals that the critical doping threshold for the onset of chaotic dynamics is exactly twice that of the magic spreading transition, $ q_c $, cf. Eq.~\eqref{eq:critical_q}:
\begin{equation}
\label{eq:otocs}
    q^\mathrm{OTOC}_c(d)\equiv -\frac{2\log(d)}{\log(\mu)} = 2 q_c(d).
\end{equation}
This indicates that generating maximally magic states requires fewer non-Clifford resources than inducing operator-level chaotic behavior.  
While magic spreading signals the breakdown of stabilizer structure at the state level, the decay of $ \mathrm{OTOC}_6 $ probes a stronger form of complexity associated with operator scrambling, which emerges only when the circuit approximates unitary designs.  
This distinction highlights the layered structure of quantum complexity and supports the existence of an intermediate regime where the system exhibits maximal magic yet remains non-chaotic. 
Moreover, this finding echoes the relationship between the nonstabilizerness of random states~\cite{turkeshi2025pauli} and the operator stabilizer entropy of random observables~\cite{dowling2024magic}.  
The generalized stabilizer purities used here naturally extend the operator stabilizer entropy to qudit systems and provide direct bounds on implementation costs for $ d \geq 3 $~\cite{rudolph2025paulipropagationcomputationalframework, angrisani2024classicallyestimatingobservablesnoiseless, lerch2024efficientquantumenhancedclassicalsimulation}, offering a concrete avenue for further exploration.

These results confirm that $ \mathrm{OTOC}_6 $ serves as a minimal and analytically tractable probe for the emergence of quantum chaos in doped Clifford circuits for $d=3$.  
A similar formula and analysis apply to higher qudit dimensions—for example, $ d = 5 $—although the expression in Eq.~\eqref{eq:otoc6d3} must be adapted to include additional terms. 
As a benchmark, in Fig.~\ref{fig:Dotoc6} we plot the relative deviation
\begin{equation}
    \Delta \mathrm{OTOC}_6 = \frac{\mathrm{OTOC}_6}{\mathrm{OTOC}_6^\mathrm{Haar}}
    \label{eq:DeltaOTOC}
\end{equation}
for $ d = 3 $ and $ d = 5 $. The results confirm that the saturation time $ N_T $ grows linearly with system size, consistent with our theoretical predictions. 
Overall, these results on the $ \mathrm{OTOC}_6 $ provide a precise and analytically tractable probe of quantum chaos in doped Clifford circuits.  
Its behavior confirms that a finite fraction of non-Clifford resources is required to drive the system into a chaotic regime and to escape classical simulability.  
These findings further highlight the power of replica techniques in diagnosing quantum complexity and the emergence of randomness in many-body dynamics.

\section{Discussion and conclusion}
Throughout this work, we have carried out a detailed analysis of the complexity properties of qudit doped Clifford circuits from multiple perspectives.  
By considering an analytically tractable architecture composed of global random Clifford gates interleaved with single-qudit non-Clifford resources, we characterized the growth of magic as a function of the number of doped layers, culminating in a transition, at a critical doping rate $q_c(d)$, to a universal regime. 
Notably, this critical point depends explicitly on the local dimension. In particular, higher-dimensional qudits exhibit a faster onset of complexity. Even more striking is the agreement with numerical simulations of doped brickwork circuits, which—despite being built from local rather than global Clifford gates—perfectly match the analytical predictions.  
This suggests a universal scaling of magic growth governed solely by the periodicity of doping, rather than circuit details. 
We also examined anticoncentration and entanglement properties, showing that a logarithmic number of non-Clifford gates $N_T\sim \log(N)$ is sufficient for the doped circuit to approximate the Haar unitary expectation to precision $\varepsilon$.
We then linked these observables to nonstabilizerness via antiflatness metrics for qutrit systems ($d=3$).
This formula is more involved than the analogues for qubit systems, and in particular, requires several GSE operators. 
Finally, we explored the emergence of quantum chaos via the out-of-time-order correlator for $ k = 3 $ replicas, which is sufficient to detect non-Clifford effects in qudits.  
We showed that an extensive number of doped layers, $ N_T = \mathcal{O}(N) $, is required to drive the OTOC to its Haar value.  
The corresponding threshold, $ q_c^\mathrm{OTOC}(d) = 2 q_c(d) $, is twice that of magic spreading. 
Consequently, higher local dimension lowers the critical doping for saturating Haar-like behavior, yet it also defines an intermediate regime in which states are maximally magical while operators remain non-chaotic. A systematic study of these regimes and their impact on the classical simulation of doped circuits is left for future work.

These results open new directions for studying resource theories in multiqudit systems, particularly in relation to complexity growth and the onset of quantum universality.  
A key question for future work is the robustness of the dynamical magic transition in the presence of noise.  
In realistic settings—especially on near-term quantum hardware—local noise may suppress or reshape the buildup of non-Clifford resources.  
It remains to be seen whether the observed dynamical transition of magic persists under noisy evolution, or if it is qualitatively altered by different error channels, including both coherent and incoherent noise.  
On a more practical ground, these insights could be extended to tensor network architectures, which provide a flexible framework for analyzing complexity and simulation cost.  
Recent work on Clifford-augmented matrix product states~\cite{fux2025disentanglingunitarydynamicsclassically, liu2024classicalsimulabilitycliffordtcircuits, huang2025augmentingdensitymatrixrenormalization, qian2024augmenting} offers a promising route to compressing quantum states with sparse non-Clifford content, and may benefit from the diagnostics developed here.

\begin{acknowledgments}
We thank N. Dowling, A. Hamma, M. Heinrich, L. Leone, P. Sierant, and E. Tirrito, for discussions. 
We acknowledge support from DFG under Germany's Excellence Strategy – Cluster of Excellence Matter and Light for Quantum Computing (ML4Q) EXC 2004/1 – 390534769, and DFG Collaborative Research Center (CRC) 183 Project No. 277101999 - project B01.

\paragraph{Code and Data Availability.} The notebook for the symbolic computation, together with the code and the data for our simulations, is publicly available at Ref.~\cite{dataset}. 
\end{acknowledgments}

\appendix 
\section{Haar unitary averages}\label{AppendixA}

We now review two key analytical results for global Haar-random evolution, which serve as theoretical benchmarks for diagnosing chaos in doped Clifford circuits.

The first concerns the computation of the generalized stabilizer entropy (GSE) under a global Haar-random unitary $ U \in U(d^N) $, as presented in Ref.~\cite{turkeshi2023measuring}.  
For $ d \geq 3 $ and $ k = 3 $, the relevant commutant operator defining the magic measure is $ \Omega_d $, given in Eq.~\eqref{eq:Omega_d}.  
Considering a Haar-random state $ \ket{\Psi} = U\ket{0}^{\otimes N} $, we evaluate the ensemble-averaged generalized purity,
\begin{equation}
    \zeta_{\Omega_d}^\mathrm{Haar} \equiv \mathbb{E}_{\mathrm{Haar}}\left[\bbra{\Omega_d} (U \otimes U^*)^{\otimes 3} \kket{0,0}^{\otimes 3} \right].
\end{equation}
By invoking Schur-Weyl duality, we recognize that the contraction of symmetric group elements $ \sigma \in S_3 $ with the initial product state yields a trivial contribution, allowing for an exact resummation over the Weingarten matrix elements.  
Using the identity $ \sum_{\sigma \in S_k} \mathrm{Wg}_{\sigma \pi}(d) = \left[\prod_{m=0}^{k-1}(d+m)\right]^{-1} $, the average reduces to
\begin{equation}
    \zeta_{\Omega_d}^\mathrm{Haar} = \frac{1}{d^N(d^N + 1)(d^N + 2)} \sum_{\sigma \in S_3} G_{\sigma \Omega_d}(d^N),
\end{equation}
where $ G_{\sigma \Omega_d}(d^N) = \mathrm{Tr}(T_\sigma \Omega_d)^N $ encodes contributions from different permutation structures.  
Evaluating the traces yields the compact result
\begin{equation}
    \zeta_{\Omega_d}^\mathrm{Haar} = \frac{3}{d^N + 2}.
\end{equation}

The second benchmark involves the Haar-averaged six-point out-of-time-ordered correlator (OTOC).  
From Eq.~\eqref{eq:otoc6_general}, the expression under Haar evolution simplifies to
\begin{equation}
    \mathrm{OTOC}_6^{\mathrm{Haar}} = \frac{1}{d^N} \sum_{\tau, \sigma \in S_3} \llangle T_{(123)}\, P \otimes P \otimes (P^2)^\dagger \,\big|\, \tau \rrangle \; \mathrm{Wg}_{\tau, \sigma}(d^N) \; \llangle \sigma \,\big|\, Q \otimes Q \otimes (Q^2)^\dagger \rrangle,
\end{equation}
where the dynamics are fully governed by the symmetric group.  
Carrying out the sum using known expressions for Weingarten functions yields the final result reported in Eq.~\eqref{eq:otocHaar}.
\bibliographystyle{quantum}
\bibliography{filtered_bib}

\end{document}